      \documentstyle[12pt,graphicx]{article}                
\begin{document}
\baselineskip 18pt
\def\today{\ifcase\month\or
 January\or February\or March\or April\or May\or June\or
 July\or August\or September\or October\or November\or December\fi
 \space\number\day, \number\year}
\def\thebibliography#1{\section*{References\markboth
 {References}{References}}\list
 {[\arabic{enumi}]}{\settowidth\labelwidth{[#1]}
 \leftmargin\labelwidth
 \advance\leftmargin\labelsep
 \usecounter{enumi}}
 \def\newblock{\hskip .11em plus .33em minus .07em}
 \sloppy
 \sfcode`\.=1000\relax}
\let\endthebibliography=\endlist
\def\lsim{\ ^<\llap{$_\sim$}\ }
\def\gsim{\ ^>\llap{$_\sim$}\ }
\def\r2{\sqrt 2}
\def\beq{\begin{equation}}
\def\eeq{\end{equation}}
\def\beqn{\begin{eqnarray}}
\def\eeqn{\end{eqnarray}}
\def\rmuu{\gamma^{\mu}}
\def\rmud{\gamma_{\mu}}
\def\PL{{1-\gamma_5\over 2}}
\def\PR{{1+\gamma_5\over 2}}
\def\sinW2{\sin^2\theta_W}
\def\AEM{\alpha_{EM}}
\def\mul{M_{\tilde{u} L}^2}
\def\mur{M_{\tilde{u} R}^2}
\def\mdl{M_{\tilde{d} L}^2}
\def\mdr{M_{\tilde{d} R}^2}
\def\mz2{M_{z}^2}
\def\c2b{\cos 2\beta}
\def\au{A_u}         
\def\ad{A_d}
\def\cob{\cot \beta}
\def\v#1{v_#1}
\def\tb{\tan\beta}
\def\epem{$e^+e^-$}
\def\KK{$K^0$-$\bar{K^0}$}
\def\wi{\omega_i}
\def\xj{\chi_j}
\def\Wmu{W_\mu}
\def\Wnu{W_\nu}
\def\m#1{{\tilde m}_#1}
\def\mH{m_H}
\def\mw#1{{\tilde m}_{\omega #1}}
\def\mx#1{{\tilde m}_{\chi^{0}_#1}}
\def\mc#1{{\tilde m}_{\chi^{+}_#1}}
\def\mwi{{\tilde m}_{\omega i}}
\def\mxi{{\tilde m}_{\chi^{0}_i}}
\def\mci{{\tilde m}_{\chi^{+}_i}}
\def\mz{M_z}
\def\sw{\sin\theta_W}
\def\cw{\cos\theta_W}
\def\cb{\cos\beta}
\def\sb{\sin\beta}
\def\rwi{r_{\omega i}}
\def\rxj{r_{\chi j}}
\def\rfp{r_f'}
\def\Kik{K_{ik}}
\def\Fq2{F_{2}(q^2)}
\def\f{\({\cal F}\)}
\def\d1{{\f(\tilde c;\tilde s;\tilde W)+ \f(\tilde c;\tilde \mu;\tilde W)}}
\def\tw{\tan\theta_W}
\def\sec2w{sec^2\theta_W}

\begin{titlepage}
\begin{flushright}
{CERN-TH/2001-047 }\\
\end{flushright}
\begin{center}
{\large {\bf  Analysis of Couplings with Large Tensor Representations 
in SO(2N) and Proton Decay}}\\
\vskip 0.5 true cm
\vspace{2cm}
\renewcommand{\thefootnote}
{\fnsymbol{footnote}}
 Pran Nath$^{a,b,c,d}$ and Raza M. Syed$^b$  
\vskip 0.5 true cm
\end{center}

\noindent
{a. Theoretical Physics Division, CERN CH-1211, Geneve 23, 
Switzerland}\\
 {b. Department of Physics, Northeastern University,
Boston, MA 02115-5000, USA\footnote{Permanent address of P.N.}} \\
{c. Physikalisches Institut, Universit\"at Bonn, 
Nussallee 12, D-53115 Bonn, Germany}\\
{d. Max-Planck Institute fuer Kernphysik, Saupfercheckweg 1, D-69117 
Heidelberg, Germany}\\ 
\vskip 1.0 true cm
\centerline{\bf Abstract}
\medskip
We develop techniques for the analysis of SO(2N) invariant couplings
which allow a full exhibition of the SU(N) invariant 
 content of the spinor and tensor
representations. The technique utilizes a basis consisting of a specific set of
reducible SU(N) tensors in terms of which the SO(2N) invariant couplings 
have a simple expansion.
The technique is specially useful for couplings 
involving large tensor representations.
We exhibit the technique by performing a complete determination of the 
trilinear
couplings in the superpotential for the case of SO(10) involving the
16  plet of matter, i.e., we give a full determination of the 
 ${ 16-16-10_s}$, ${ 16-16-120_a}$ and ${ 16-16-\bar{126}_s}$
  couplings. The possible role of large tensor representations
   in the generation of quark lepton textures is discussed.
   It is shown that the couplings involving $\bar{126}$ dimensional
    representation generate extra zeros in the Higgs triplet textures
    which can lead to an enhancement of the proton decay lifetime
    by a factor of $10^3$. These results also have implications
    for neutrino mass textures.  
\end{titlepage}

\section{Introduction}
The group SO(10) is one of the candidates for a theory of grand unification
and  has come under increasing scrutiny since the early work of
Ref.\cite{georgi} because of its desirable features such as
the unification of  one generation of quarks and leptons in a
single multiplet and a relatively natural way in which the 
doublet-triplet splitting can be achieved in the model\cite{doubtrip}.
 On the technical side the introduction of the oscillator
technique\cite{sakita,wilczek,nandi} in SO(10) analyses has proven useful.
However, SO(10) matter interactions may involve large tensor 
representations, i.e., 120, $\bar{126}$ and 210. Specifically the
representations $120$ and $\bar{126}$ have already surfaced in the 
analyses of quark, charged lepton and neutrino mass 
textures\cite{gn,harvey,babu,largerep,mahan}. However, 
 a full analysis of the couplings
of such large representations such as $16-16-120$ and $16-16-\bar{126}$ 
does not exist in the literature in any explicit form.   
  We develop here a systematic approach that enables one to 
carry out a full computation of such couplings with relative
ease. We then illustrate our technique by giving a complete analysis
of the trilinear superpotential with the 16  plet of matter.
Since $16\times 16= 10_s+ 120_a + 126_s$ we give an explicit 
computation of the $16-16-10_s$, $16-16-120_a$, and $16-16-\bar{126}_s$
couplings.

Our technique is a natural extension of the work of 
Refs.\cite{sakita,wilczek} which introduced 
the oscillator expansion in the analysis of SO(2N) 
interactions [One may also use completely group theoretic
methods to compute the couplings as done in $E_6$ model
building analysis of Ref.\cite{anderson}.  Our technique is field
theoretic and more straightforward.].
We briefly review this analysis first. In the oscillator technique of
Refs.\cite{sakita} one defines a set of N operators
$b_i$ (i=1,...,N) obeying the anti-commutation rules
\begin{equation}
\{b_i,b_j^{\dagger}\}=\delta_{ij}; ~~~\{b_i,b_j\}=0
\end{equation}
and represents the set of 2N operators $\Gamma_{\mu}$ 
($\mu=1,2,..,2N$) by 
\begin{equation}
\Gamma_{2i}= (b_i+ b_i^{\dagger});~~~ 
\Gamma_{2i-1}= -i(b_i- b_i^{\dagger})
\end{equation}
where $\Gamma_{\mu}$ satisfy a rank 2N Clifford algebra
$\{\Gamma_{\mu},\Gamma_{\nu}\}=2\delta_{\mu\nu}$. 
The group SO(2N) has a $2^N$ dimensional spinor representation $\psi$.
This representation can be split into $2^{N-1}$ dimensional representation
under the action of the chirality operator so that
\begin{equation}
\psi_{\pm}=\frac{1}{2}(1 \pm \Gamma_0)\psi,~~~
\Gamma_0=i^N  \Gamma_1\Gamma_2...\Gamma_{2N}
\end{equation}
where $\psi_{\pm}$ are each $2^{N-1}$ dimensional. 
In the analysis of SO(2N) interactions one encounters couplings of the
type $\tilde\psi B\Gamma_{\mu}..\Gamma_{\sigma}\psi
\phi_{\mu ..\sigma}$ where  
$B (=\prod_{\mu =odd}\Gamma_{\mu})$ is an SO(2N) charge conjugation 
matrix. We wish to develop here a simple technique for the explicit 
evaluation of the couplings in terms of the physical degrees of freedom
even for the  case when the tensor representation that couples has 
a large dimensionality.  

\section{The Basic Theorem}
We begin with the observation that the natural basis for the expansion of the
SO(2N) vertex is in terms of a specific set of SU(N) reducible 
tensors which we define below. 
We introduce the notation  
$\phi_{c_i}=\phi_{2i}+i\phi_{2i-1}$ and
$\phi_{\bar c_i}=\phi_{2i}-i\phi_{2i-1}$ which can be extended immediately
 to define the quantity $\phi_{c_ic_j\bar c_k..}$
with an arbitrary number of unbarred and barred indices where each 
 c index can be expanded out so that  
 $\phi_{c_ic_j\bar c_k..}=\phi_{2ic_j\bar c_k...}+i\phi_{2i-1c_j\bar c_k..}$,
 $\phi_{c_ic_j\bar c_k..}=\phi_{c_ic_j2k...}-i\phi_{c_ic_j2k-1...}$,etc.
Thus, for example, the quantity  $\phi_{c_ic_j\bar c_k...c_N}$ is a sum of 
 $2^N$ terms gotten by expanding all the c indices.  
$\phi_{c_ic_j\bar c_k...c_n}$ is completely anti-symmetric 
in the interchange of its c indices whether unbarred or barred.
We now make the observation that the object 
$\phi_{c_ic_j\bar c_k...c_n}$ transforms like a reducible representation
of SU(N). Thus if we are able to compute the  SO(2N) invariant couplings 
 in terms of these reducible tensors of SU(N) then
there remains only the further step of decomposing the reducible 
 tensors into their irreducible parts. Finally, one can
take the result obtained in terms of the SU(N) irreducible 
representations and expand out in terms of the particles
of the model. 

   The result essential to our analysis is the theorem 
that the quantity 
$\Gamma_{\mu}\Gamma_{\nu}\Gamma_{\lambda}..\Gamma_{\sigma}$
$\phi_{\mu\nu\lambda ..\sigma}$ can be expanded in the following form 
\begin{eqnarray}
\Gamma_{\mu}\Gamma_{\nu}\Gamma_{\lambda}..\Gamma_{\sigma}
\phi_{\mu\nu\lambda ..\sigma}= b_i^{\dagger} b_j^{\dagger}b_k^{\dagger}
..b_n^{\dagger} \phi_{c_ic_jc_k...c_n}+
(b_i b_j^{\dagger}b_k^{\dagger}
..b_n^{\dagger} \phi_{\bar c_ic_jc_k...c_n}+~perms)\nonumber\\
+(b_i b_jb_k^{\dagger}..b_n^{\dagger} \phi_{\bar c_i\bar c_jc_{k}..c_n}
+~perms)+ ....+(b_ib_jb_k..b_{n-1}b_n^{\dagger} 
\phi_{\bar c_i\bar c_j\bar c_k..\bar c_{n-1}c_n} +~perms)+\nonumber\\
+ b_ib_jb_k.....b_n \phi_{\bar c_i\bar c_j\bar c_k...\bar c_n}
\end{eqnarray}
 Eq.(4) is  
 the basic result we need in the analysis of the SO(2N) invariant 
couplings. It is found convenient to arrange the right hand side of 
Eq.(4)  in a normal ordered form by which we mean that all the $b$'s are 
either to the
right or to the left and all the $b^{\dagger}$'s are either to the left
or to the right  using strictly
the anti-commutation relations on the $b$'s and $b^{\dagger}$'s
of Eq.(1). When a pair of $b^{\dagger}$ and b have a summed index
such as $b_n^{\dagger}b_n$ we will move them together either to 
the left or to the right.
 After normal ordering one decomposes $\phi_{c_ic_j\bar c_k..c_n}$
into its SU(N) irreducible components. 
The final step consists of carrying out 
 the process of removing all the $b$ and $b^{\dagger}$ using the 
anti-commutation relation Eq.(1) along with the condition
$b_i|0>=0$ which allows us to compute the
couplings in the SU(N) invariant decomposition.

\section{The 120 plet and the $\bar {126}$ plet tensor couplings}
The above procedure makes it straightforward to analyze the 
 SO(2N) invariant couplings involving large tensor representations.
 As an illustration of our technique we give  a complete 
determination of the superpotential at the trilinear level involving 
two spinor representations which consists of the couplings 
${ 16\times 16\times 10}$, ${ 16\times 16\times 120}$
and ${ 16\times 16\times \bar {126}}$.
We begin by computing
 the $ {16\times 16\times 10}$ coupling which is given by
 \begin{equation}
       f_{ab}\tilde \psi_a B \Gamma_{\mu}\psi_b\phi_{\mu}
\end{equation}
where a, b are the generation indices.
 Following the procedure of  sec.2 we decompose the vertex 
 so that 
 \begin{equation}
\Gamma_{\mu}\phi_{\mu}= b_i\phi_{\bar c_i}+b_i^{\dagger} \phi_{c_i}
\end{equation} 
In Eq.(6) the tensors are already in their irreducible form and one can 
identify $\phi_{c_i}$ with the 5 plet of Higgs and $\phi_{\bar c_i}$
with the $\bar 5$ plet of Higgs.  To normalize the tensors we
define  
$ H_{1i}=\frac{1}{\sqrt 2} \phi_{\bar c_i}$,
$H_{2}^i=\frac{1}{\sqrt 2}\phi_{c_i}$, 
so that the kinetic energy 
$-\partial_{\alpha}\phi_{\mu}\partial^{\alpha}\phi_{\mu}^{\dagger}$  
of the tensor $\phi_{\mu}$ takes the form 
 $-\partial_{\alpha}H_{1i}\partial^{\alpha} H_{1i}^{\dagger}$
 $-\partial_{\alpha}H_{2}^i\partial^{\alpha} H_{2}^{i\dagger} $.
 For the computation of the superpotential we need to 
 expand the 16 plet spinor representation
  $\psi_+$ in its oscillator modes
 \begin{equation}
|\psi_+>= |0>M_0 +\frac{1}{2}b_i^{\dagger} b_j^{\dagger}|0>M^{ij}+  
\frac{1}{24}\epsilon^{ijklm}b_i^{\dagger} b_j^{\dagger}
b_k^{\dagger} b_l^{\dagger}|0>M_{m}'
\end{equation}  
so that $\psi_+$ contains $1_M+10_M+\bar 5_M$ in its SU(5) decomposition.
 Using the above  we compute the $16-16-10$ couplings and find
 \begin{eqnarray}
  W^{(10)}=(2\sqrt 2 i)f^{(+)}_{ab}(M^{ij}_aM'_{ib}H_{1j}
  -M_{0a}M'_{ib}H_2^i + \frac{1}{8}\epsilon_{ijklm}M^{ij}_aM^{kl}_b H_2^m)
\end{eqnarray} 
where $f_{ab}^{(\pm)}$ are defined by
\begin{equation}
f_{ab}^{(\pm)}=\frac{1}{2} (f_{ab} \pm f_{ba})
\end{equation}
and $f_{ab}^{(\pm)}$ are symmetric (antisymmetric)
under the interchange of generation indices a and b.
As expected the 16-16-10 couplings given by Eq.(8) are 
correctly symmetric in the generation indices.  We note that the 
  couplings have the  SU(5) invariant structure consisting of 
  $1_M-\bar 5_M-5_H$,
  $10_M-\bar 5_M-\bar 5_H$ and $10_M -10_M-5_H$.
   
Next we discuss the 16-16-120 coupling which is given by
 \begin{equation}
 \frac{1}{3!}
f_{ab}\tilde \psi_a B \Gamma_{\mu}\Gamma_{\nu}\Gamma_{\lambda}
\psi_b\phi_{\mu\nu\lambda}
\end{equation}
 We expand the vertex using Eq.(4) and find
  \begin{eqnarray}
\Gamma_{\mu}\Gamma_{\nu}\Gamma_{\lambda}\phi_{\mu\nu\lambda}
= b_ib_jb_k \phi_{\bar c_i\bar c_j\bar c_k} 
+ b_i^{\dagger}b_j^{\dagger}b_k^{\dagger} \phi_{c_i c_j c_k}
+ 3 (b_i^{\dagger}b_jb_k \phi_{c_i\bar c_j \bar c_k}+
 b_i^{\dagger}b_j^{\dagger}b_k \phi_{c_i c_j \bar c_k}) \nonumber\\
+(3 b_i \phi_{\bar c_n c_n \bar c_i}+ 3b_i^{\dagger} 
\phi_{\bar c_n c_n c_i} )
\end{eqnarray}
The 120 plet of SO(10) has the SU(5) decomposition 
$120=5+\bar 5+10+\bar{10}+45+\bar{45}$. Eq.(11) can be decomposed in
terms of these irreducible SU(5) tensors as explained in the
appendix. A straightforward computation using 
Eq.(11)\footnote{The symmetrical arrangement in the first brace
of Eq.(11) is necessary for achieving an automatic anti-symmetry
in the generation indices for the SU(5) $10_M-10_M-45_H$ coupling.}
 and the normalization of Eq.(23) in the appendix gives  
 \begin{eqnarray}
W^{(120)}=i\frac{2}{\sqrt 3}f_{ab}^{(-)}(2 M_{0a} M_{ib}h^i +
M^{ij}_aM_{0b}h_{ij}
+M_{ia}M_{jb}h^{ij}\nonumber\\
-M^{ij}_aM_{ib}h_j+  M_{ia}M_b^{jk}h^i_{jk}
-\frac{1}{4}\epsilon_{ijklm}M_a^{ij}M_b^{mn}h^{kl}_n)  
\end{eqnarray}
The front factor $f_{ab}^{(-)}$ in Eq.(12) exhibits 
 correctly the  anti-symmetry in the generation indices.
Further, the couplings have the 
$1_M-\bar 5_M-5_H$, $1_M-10_M-\bar{10}_H$, 
$\bar 5_M-\bar 5_M-10_H$, $10_M-\bar 5_M-\bar 5_H$, 
$\bar 5_M-10_M-\bar{45}_H$ and $10_M-10_M-45_H$
SU(5) invariant structures.

  We now turn to the most difficult of the three cases, 
  i.e., the ${ 16-16-\bar {126}}$ coupling which is given by
\begin{equation}
 \frac{1}{5!}f_{ab}\tilde \psi_a B 
\Gamma_{\mu}\Gamma_{\nu}\Gamma_{\lambda}
\Gamma_{\rho}\Gamma_{\sigma}\psi_b\Delta_{\mu\nu\lambda\rho\sigma}
\end{equation}
where  
$\Delta_{\mu\nu\lambda\rho\sigma}$ is  252 dimensional and 
can be decomposed so that 
$\Delta_{\mu\nu\lambda\rho\sigma}$=$\bar \phi_{\mu\nu\lambda\rho\sigma}$+
$\phi_{\mu\nu\lambda\rho\sigma}$, where\cite{harvey}
\begin{equation}
\left(\matrix{\bar\phi_{\mu\nu\lambda\rho\sigma}\cr
 \phi_{\mu\nu\lambda\rho\sigma}}\right)=
   \frac{1}{2}(\delta_{\mu\alpha}
\delta_{\nu\beta}\delta_{\rho\gamma}\delta_{\lambda\delta}\delta_{\sigma\theta}
\pm \frac{i}{5!}\epsilon_{\mu\nu\rho\lambda\sigma\alpha\beta\gamma\delta\theta})
\Delta_{\alpha\beta\gamma\delta\theta}
\end{equation}
and where the $\bar\phi_{\mu\nu\lambda\rho\sigma}$ is the ${\bar{126}}$ plet
 and $\phi_{\mu\nu\lambda\rho\sigma}$ is the ${ 126}$ plet representation.
 It is only the ${\bar{126}}$ that couples in Eq.(13) with the 
 ${ 16}$ plet 
 spinors. However, for the reduction of the SO(10) vertex it is more
 convenient initially to work with the full 252 dimensional tensor 
 and in the final computation  only the ${\bar{126}}$ couplings will survive. 
 We begin by expanding
$\Gamma_{\mu}\Gamma_{\nu}\Gamma_{\lambda}
\Gamma_{\rho}\Gamma_{\sigma}\Delta_{\mu\nu\lambda\rho\sigma}$
 using Eq.(4) following steps similar to the previous case using 
  normal ordering  and further 
  decomposing the tensors into their irreducible components.
The $\bar{126}$ and the 126 dimensional representations break 
into the SU(5) irreducible parts as ${\bar{126}=1+5+
\bar {10}+ 15+\bar{45}+50}$ and 
${ 126=1+\bar 5+10+ \bar{15}+ 45 +\bar{50}}$.
The details of the decomposition are given in the appendix. 
Using Eq.(25) in the appendix  a straightforward 
analysis gives 
\begin{eqnarray}
W^{(\bar{126})}= if_{ab}^{(+)} {\frac{\sqrt 2}{\sqrt {15}}[-\sqrt 2}
M_{0a} M_{0b}h-\sqrt 3  M_{0a} M'_{ib}h^i 
+M_{0a}M^{ij}_bh_{ij}\nonumber\\
-\frac{1}{8\sqrt 3}M_a^{ij}M_b^{kl}h^m\epsilon_{ijklm} 
-h^{ij}_S M'_{ia}M'_{jb}+M^{ij}_aM_{bk}'h^k_{ij}
-\frac{1}{12\sqrt 2}\epsilon_{ijklm}M_a^{lm}M_b^{rs}h^{ijk}_{rs}]
\end{eqnarray}
where we have expressed the result in terms of
fields $h^i, h_{ij}$ etc with normalizations given by  
 Eq.(27).
As in the ${10}$ plet tensor case the couplings are symmetric under 
the interchange of generation
indices. Further, the ${16-16-\bar {126}}$ coupling has the SU(5) structure
consisting of ${ 1_M-1_M-1_H}$, 
${ 1_M-\bar 5_M-5_H}$, ${ 1_M-10_M-\bar{10}_H}$,
${ 10_M-10_M-5_H}$, ${ \bar 5_M-\bar 5_M-15_H}$,  
${ 10_M-\bar 5_M-\bar{45}_H}$ and $10_M-10_M-50_M$. 
A similar analysis can be carried out
for the tensor couplings involving $\bar{16}-16$ which includes
$\bar{16}-16-45$ and $\bar{16}-16-210$ couplings. These will be discussed
elsewhere.

\section{Large representations, textures and proton lifetime}
Proton decay is an important signal for grand unification and
detailed analyses for the proton lifetime exist in SU(5) 
models\cite{ellis,nac,in}  and in
SO(10) models\cite{lucas,bpw}.
 In this section we discuss the possibility that couplings in the
 superpotential that involve large representations can drastically 
 change the Higgs triplet textures and affect proton decay in
  a very significant manner. 
These results are of significance in view of the recent 
data from SuperKamiokande which has significantly improved
the limit on the proton decay mode $p\rightarrow \nu+K^+$.
Thus the most recent limit from SuperKamiokande gives 
$\tau/B(p\rightarrow \bar \nu+K^+)> 1.9 \times 10^{33}$ yr\cite{totsuka}. 
 At the same time
there is a new lattice gauge evaluation of the three quark matrix
elements $\alpha$ and $\beta$ of the nucleon wave function\cite{aoki}
(where $\alpha$ and $\beta$ are defined by\cite{brodsky} 
$\epsilon_{abc}<0|\epsilon_{\alpha\beta}
d^{\alpha}_{aR}u^{\beta}_{bR}u^{\gamma}_{cL}|p>$=$\alpha u^{\gamma}_L$
and
$\epsilon_{abc}<0|\epsilon_{\alpha\beta}
d^{\alpha}_{aL}u^{\beta}_{bL}u^{\gamma}_{cL}|p>$=$\beta u^{\gamma}_L$).
Previous evaluations of these quantities have varied over a wide
range from $\beta =0.003GeV^3$\cite{donoghue} to 
$\beta =0.03GeV^3$\cite{brodsky,hara} while recent 
p decay analyses have often used the lattice gauge evaluation of
Gavela et.al.\cite{gavela} which gives
 $\beta =(5.6\pm0.5)\times 10^{-3}GeV^3$.
 The more recent evaluation of Ref.\cite{aoki} gives
  $\alpha =-0.015(1) GeV^3$ and 
$\beta =0.014(1) GeV^3$ which is a factor of about two and a half times
 larger than
the evaluation of Ref.\cite{gavela}.
  The new experimental limit on the proton decay lifetime\cite{totsuka} 
  combined with the new lattice gauge
  evaluations have begun to constrain
the SUSY GUT models prompting some reanalyses\cite{dmr,altarelli}.
In this context the enhancement of the proton lifetime by textures
is of interest. 
To make this idea more concrete we define textures in the low 
energy theory in the quark lepton sector of the theory
just below the GUT scale as follows  
\beqn
W_Y= -M_HH_{1t}H_{2t}+ (l A^E e^c h_1 + q A^D d^c h_1 +h_2u^cA^Uq)\nonumber\\ 
+ (qB^El H_{1t}+ \epsilon_{abc}H_{1ta}d^c_bB^Du^c_c
+ H_{2ta}u^c_aB^Ue^c 
+\epsilon_{abc}H_{2t}^au_bC^Ud_c) 
\eeqn
where $A^E$, $A^D$ and $A^U$  are the textures in the
Higgs doublet sector and $B^E, B^D$, $B^U$ and $C^U$ are the textures 
in the Higgs triplet sector. A classification of the
possible textures in the Higgs doublet sector is given in 
Refs.\cite{gj,ross}.
For our purpose here we adopt the textures in the Higgs doublet sector
in the form 
\beq
A^E=
\left(\matrix{0 & f & 0 \cr
f & -3e & 0\cr
0 & 0  & d}\right),
A^D=
\left(\matrix{0 & fe^{i\phi} & 0 \cr
fe^{-i\phi} & e & 0\cr
0 & 0  & d}\right),
A^U=
\left(\matrix{0 & c & 0 \cr
c & 0 & b\cr
0 & b  & a}\right)
\eeq
As is well known\cite{gj} the appearance of -3 vs 1 in the 22 element
of $A^E$ vs $A^D$ is one of the important ingredients in achieving  
the desired quark and lepton mass hierarchy and may provide an insight 
into the nature of the fundamental coupling. 
Now the textures in the Higgs triplet sector are generally different
than those in the Higgs doublet sector and they are sensitively
dependent on the nature of GUT and  Planck scale 
physics\cite{pn}. 
The current 
experimental constraints on the proton lifetime leads us to 
conjecture that the Higgs triplet sector contains additional 
texture zeros over and above the texture zeros that appear in the
Higgs doublet sector and the coupling of the $\bar{126}$ tensor
field plays an important role in this regard. In the following we 
shall assume CP invariance and set the phases to zero in Eq.(17).
Since in this case the textures of Eq.(17) are symmetric 
it is only the 10 plet
and the $\bar{126}$ plet of Higgs couplings that enter in the
analysis and the 120 plet couplings do not. 
To exhibit the above phenomenon more concretely 
we consider on  phenomenological grounds 
 a superpotential in the Yukawa sector of the following type
\begin{eqnarray}
W_Y= f_{ij}^{(0)}(Y,M) \psi_i\psi_j \phi^{(0)}_{\bar{126}}
+f^{(d)}_{12}(Y, M)\psi_1\psi_2\phi^{(1)}_{10}
+f^{(u)}_{12}(Y, M)\psi_1\psi_2\phi^{(2)}_{10}\nonumber\\
+f^{(d)}_{22}(Y, M)\psi_2\psi_2\phi^{(1)}_{\bar{126}}
  + f^{(u)}_{23}(Y, M)\psi_2\psi_3\phi^{(2)}_{\bar{126}}
+f^{(d)}_{33}(Y, M)\psi_3\psi_3\phi^{(1)}_{10}\nonumber\\
+f^{(u)}_{33}(Y, M)\psi_3\psi_3\phi^{(2)}_{\bar{126}}
\end{eqnarray}
where $M$ is a superheavy scale
and $f^{(d)}_{ij}(Y, M)$ and $f^{(u)}_{ij}(Y, M)$ are functions
of a set of scalar fields Y which develop VEVs 
and the appropriate factors of $<Y^n\phi>/M^n$ generate the right
sizes. The model of  Eq.(18) is of the generic type discussed
in refs.\cite{harvey,mahan}.  
 We do not go into detail here regarding 
the symmetry breaking mechanism, the doublet-triplet splitting and
the mass generation for the pseudo-goldstone 
bosons. All of these topics 
have been dealt with
at some length in the previous literature\cite{doubtrip,harvey,babu,largerep}. 
Further, while models with large representations are not asymptotically
free and lead rapidly to non-perturbative physics above the 
unification scale, the effective theories below the unification scale
gotten by integration over the heavy modes are nonetheless perfectly
normal and thus such theories are acceptable unified theories.
For our purpose  here we assume 
a pattern of VEV formation for the neutral components of the Higgs 
so that $ <\phi^{(0)}_{\bar{126}}>$ develops a  VEV along the SU(5) singlet
direction (this corresponds to h in Eq.(15) developing a VEV),
 $ <\phi^{(1)}_{10}>$ develops a VEV in the $\bar 5$ plet of 
SU(5) direction
(this corresponds to $H_1$ developing a VEV in Eq.(8)),  
 $ <\phi^{(2)}_{10}>$ develops a VEV in the 5 plet  direction
 (this corresponds to $H_2$  developing a VEV in Eq.(8)),
 $<\phi^{(1)}_{\bar{126}}>$ develops a VEV  in the direction of  
 $\bar{45}$ plet of Higgs (this corresponds to $h^k_{ij}$ in Eq.(15) developing 
 a VEV), and 
  $<\phi^{(2)}_{\bar{126}}>$ develops a VEV  in the direction of  
 5 plet of Higgs (this correspons to $h^i$ in Eq.(15) developing 
 a VEV).  It is the VEV of the 45 plet that leads to -3 and 1 
 factors in $A^E$ vs $A^D$.
The superpotential of Eq.(18) with the above VEV alignments then leads
automatically to the textures in the Higgs doublet sector
of Eq.(17). 
 One may now compute the textures in the Higgs triplet sector that
 result from  superpotential of Eq.(18). One finds 
\beq
B^E=
\left(\matrix{0 & f & 0 \cr
 f & 0 & 0\cr
0 & 0  & d}\right), 
B^U=
\left(\matrix{0 & c & 0 \cr
c & 0 & 0\cr
0 & 0 & 0}\right)
\eeq
and $B^D=B^E$ and $C^U=B^U$. 
We note the existence of the additional zeros in $B^E$ and $B^D$ 
relative to $A^E$ and $A^D$ and in $B^U$ and $C^U$ relative to
$A^U$. 
We shall show shortly that the existence of the additional zeros in
$B^E$, $B^D$,$B^U$ and $C^U$ increases in a very significant manner the
proton decay lifetime. Before we discuss this enhancement in greater
detail, we wish to discuss the origin of the additional zeros.
 It is easy to see that a coupling of the matter sector with 
the $\bar {126}$ of Higgs which contributes a non-vanishing 
element in the Higgs doublet sector produces a vanishing contribution
in the lepton and baryon number violating dimension five operator
or equivalently generates a corresponding zero in the texture in
the Higgs triplet sector.
The reason for this is rather straightforward. While 
one also needs a $126$ plet of Higgs to cancel the D term generated
by the VEV of the $\bar{126}$  of Higgs, 
the 126 plet of Higgs has
no coupling with the ordinary 16 plet of 
matter. Since the only bilinear with the $\bar{126}$ in
the superpotential is of the form $126\times\bar{126}$
(i.e., one cannot write a $(126)^2$ term in the superpotential)  
one finds that 
 no lepton and baryon number violating dimension five operators arise  
as a consequence of integrating out the $126$ and $\bar{126}$
of Higgs which effectively corresponds to a texture zero in the Higgs triplet
sector. Of course, the extra zeros in the Higgs triplet sector could
also arise from accidental cancellations. However, the group theoretic
origin is more appealing.

The extra zeros in the textures in the Higgs triplet sector lead
to a substantial enhancement of the proton decay lifetime.
To see their effect we begin by integrating
out the Higgs triplet field in Eq.(16)  which generates lepton and
baryon number violating dimension five operators with the chiral
structure LLLL and RRRR.  Of these the LLLL operator involves the
textures $B^E$ and $C^U$ while the RRRR operator involves the textures
$B^D$ and $B^U$. Since the number of extra zeros in $B^E$ and $B^D$
are the same and the same holds for $B^U$ and $C^U$, it suffices to
discuss only one of these operators, and in the following we focus
on the LLLL dimension five operator. 
Here  the texture zero in the 22 element of  $B^E$ 
suppresses the $\bar\nu_{\mu} K^+$ decay mode 
of the proton by a factor
$m_d/m_s$ making the $\bar\nu_{\tau}K^+$ the dominant mode.
Since the decay channel $\bar\nu_{\mu}K^+$ is highly suppressed 
(while the decay channel 
 $\bar\nu_{e} K^+$ which is normally suppressed remains suppressed)
  we estimate that there is an over all suppression 
in all the neutrino decay channels 
from the extra zero in $B^E$ to be about a factor of about  2.
  The texture zeros in  $C^U$ lead roughly to a replacement of $m_c$ by 
$\sqrt{m_um_c}$ and
thus lead to a suppression of the proton decay lifetime
roughly  by $ m_u/m_c$.  
A similar suppression holds for decays via the RRRR dimension
five operator. Including the suppression from both $B^E$ and $C^U$ and
using  $m_c=1.35$ GeV and the up quark 
mass in the range 1-5 MeV,  one finds that the texture zeros can
lead to an enhancement of the proton decay lifetime in the 
$\bar\nu K^+$ mode by a factor
of $(1.5\pm 1)\times 10^3$ over the minimal SU(5) model. 
Such lifetimes fall in the interesting range
for the next generation of proton decay experiments. 
The texture effects are generic and similar
 effects are expected in other decay modes as well. 
The superpotential of Eq.(18) also generates  a Dirac neutrino mass
matrix which is given by 
\beq
M_{\nu LR}=
\left(\matrix{0 & c & 0 \cr
 c & 0 & -3b\cr
0 & -3b  & -3a}\right)h_2
\eeq
However, a full analysis of neutrino masses requires a model
for the Majorana mass matrix $M_{Maj}$ to generate a 
see-saw mechanism\cite{seesaw}
so that $M_{\nu LL}=-M_{\nu LR}^{T}M^{-1}_{Maj}M_{\nu LR}$
where the mass scale associated with $ M_{Maj}$ is much larger
than the mass scale that appears in $M_{\nu LR}$.
$ M_{Maj}$ depends on $f_{ij}^{(0)}$ in Eq.(18) which are 
in general additional arbitrary parameters.
[For a sample of recent analyses on neutrino masses in SO(10) see
Ref.\cite{mahan,barbi} and for a review  and a classification of models of 
neutrino masses see Ref.\cite{dorsner}].
The appearance of $M_{Maj}$ in the analysis of neutrino masses
with additional arbitrary parameters and a scale much larger
than the mass scale that appears in $M_{\nu LR}$
implies that there is not a rigid relationship between proton decay 
and neutrino masses in SO(10). Nonetheless, it is interesting to investigate
the correlation that exists between these two important 
phenomena in specific
models. 
 A more extensive analysis of this topic involving the
 detailed coupling structure of Eqs.(8), (12) and (15) as well as other
 texture possibilities in the Higgs doublet sector\cite{ross} will 
 be discussed elsewhere.
 
In conclusion, we have developed a technique 
which allows the explicit computation of the SO(2N) invariant couplings
in terms of SU(N) invariant couplings. The technique is specially useful
in the analysis of couplings involving large tensor representations.
We have illustrated the technique by carrying out a complete 
analysis of the SO(10) invariant superpotential at the trilinear level
involving interactions of matter with Higgs which consists of
 the ${ 16_a16_b10_H}$, ${ 16_a16_b120_H}$, and the 
 ${ 16_a16_b \bar{126}_H}$ couplings.
 The technique can be used
with relative ease to compute other couplings involving large
tensor representations such as ${\bar{16}-16-210}$. We note that 
the decomposition of SO(10) into multiplets of  SU(5)
is merely a convenient device for expanding the SO(10) interaction
in a compact form and does not necessarily imply a preference for 
the symmetry breaking
pattern. Indeed one can compute the SO(10) interactions using the
technique used here and then use any symmetry breaking scheme one
wishes to get to the low energy theory. We also discussed  in this
paper the phenomena that the coupling of the $\bar{126}$ with matter
leads to extra zeros in the Higgs triplet sector and the existence
of such zeros can enhance the proton decay lifetime by as much
as $10^3$.  Thus $\bar{126}$ couplings might help relieve the
constraint on SUSY GUT models because of the recent SuperKamiokande
data and improved lattice gauge calculations of $\alpha$ and $\beta$.
 The coupling involving large tensor representation given 
in sec.3 also have implications for neutrino mass textures.
Finally, the technique discussed here is 
easily extendible to models with SO(2N+1) invariance.
[Note added: The enhancement of the proton lifetime by a factor 
of $O(10^3)$ discussed here may be needed to overcome the light
sparticle spectrum (see e.g., U. Chattopadhyay and P. Nath, 
hep-ph/01021577) implied by the Brookhaven g-2 experiment,
H.N. Brown et.al, hep-ex/0102017.]

\section*{Acknowledgements}
We wish to thank Pierre Ramond for a discussion. 
One of us (PN) wishes to thank the Physics Institute at the University of Bonn
and the Max Planck Institute, Heidelberg, for hospitality
and acknowledges support from the Alexander von Humboldt Foundation.
This research was supported in part by NSF grant PHY-9901057.\\

\noindent
{\bf Appendix: Details of 120 plet and $\bar{126}$ plet couplings and
 normalizations}\\
 We discuss now the details of the 120 plet and $\bar{126}$ plet couplings.
 For the tensors that appears in the 120 plet coupling, can be 
 decomposed in their irreducible forms by the decomposition
 \begin{eqnarray}
\phi_{c_ic_j\bar c_k}=f^{ij}_k+\frac{1}{4}(\delta^i_kf^j-
\delta^j_kf^i),~~~
\phi_{c_i\bar c_j\bar c_k}=f^{i}_{jk}+\frac{1}{4}(\delta^i_jf_k-
\delta^i_kf_j)\nonumber\\
\phi_{c_ic_jc_k}=\epsilon^{ijklm}f_{lm},~~~  
\phi_{\bar c_i\bar c_j\bar c_k}=\epsilon_{ijklm}f^{lm},
~~~\phi_{\bar c_nc_n c_i}=f^i, \phi_{\bar c_nc_n \bar c_i}=f_i
\end{eqnarray} 
 where $f^{ij}_k$ and $f^i_{jk}$ are traceless and are the 45 plet and the
 $\bar{45}$ plet representations of SU(5).
 The irreducible tensors $f^i, f^{ij}$ etc are not yet properly normalized.
 To normalize them we make the following redefinition of fields
 \begin{eqnarray}
f^{i}=\frac{4}{\sqrt 3} h^{i},~~~
f^{ij}=\frac{1}{\sqrt 3} h^{ij},~~~
f^{ij}_k=\frac{2}{\sqrt 3} h^{ij}_k\nonumber\\
f_{i}=\frac{4}{\sqrt 3} h_{i},~~~
f_{ij}=\frac{1}{\sqrt 3} h_{ij},~~
f^{i}_{jk}=\frac{2}{\sqrt 3} h^{i}_{jk}
\end{eqnarray}
In terms of the redefined fields the kinetic energy term for the 120 
multiplet which is given by $-\partial_{\alpha}\phi_{\mu\nu\lambda}$
$\partial^{\alpha}\phi_{\mu\nu\lambda}^{\dagger}$
takes on the form 
\begin{eqnarray}
L_{kin}^{(120)}=-(\frac{1}{2}\partial_{\alpha} h^{ij}
\partial^{\alpha} h^{ij\dagger} 
+\frac{1}{2}\partial_{\alpha} h_{ij}
\partial^{\alpha} h_{ij}^{\dagger} 
+\frac{1}{2}\partial_{\alpha} h^{ij}_k
\partial^{\alpha} h^{ij\dagger}_k  \nonumber\\
 +\frac{1}{2}\partial_{\alpha} h_{jk}^i
\partial^{\alpha}h_{jk}^{i\dagger} 
+\partial_{\alpha} h^{i}
\partial^{\alpha} h^{i\dagger}
+\partial_{\alpha} h_{i}
\partial^{\alpha} h_{i}^{\dagger}) 
\end{eqnarray}
where the factors of 1/2 are to account for (ij) permutations.

Next we discuss details of the $\bar{126}$ plet couplings.  Here the 
reducible tensors that enter in the expansion of the vertex are
$\Delta_{c_ic_jc_kc_l\bar c_m}$,  $\Delta_{c_ic_jc_k\bar c_l\bar c_m}$
etc. These can be decomposed into their irreducible parts as follows
\begin{eqnarray}
\Delta_{c_ic_jc_kc_l\bar c_m}=f^{ijkl}_m+\frac{1}{2}
(\delta^i_mf^{jkl}-\delta^j_mf^{ikl}+ 
\delta^k_mf^{ijl}-\delta^l_mf^{ijk} )\nonumber\\
\Delta_{c_ic_jc_k\bar c_l\bar c_m}=f^{ijk}_{lm}+\frac{1}{2}
(\delta^i_lf^{jk}_m -\delta^j_lf^{ik}_m
+\delta^k_lf^{ij}_m -\delta^i_mf^{jk}_l
+\delta^j_mf^{ik}_l -\delta^k_mf^{ij}_l)\nonumber\\
+\frac{1}{12}(\delta^i_l\delta^j_mf^k-\delta^j_l\delta^i_mf^k
-\delta^i_l\delta^k_mf^j+\delta^k_l\delta^i_mf^j
+\delta^j_l\delta^k_mf^i-\delta^k_l\delta^j_mf^i)\nonumber\\
\Delta_{c_ic_j\bar c_k\bar c_l\bar c_m}=f^{ij}_{klm}+\frac{1}{2}
(\delta^i_kf^{j}_{lm} -\delta^i_lf^{j}_{km}
+\delta^i_mf^{j}_{kl} -\delta^j_kf^{i}_{lm}
+\delta^j_lf^{i}_{km} -\delta^j_mf^{i}_{kl})\nonumber\\
+\frac{1}{12}(\delta^i_k\delta^j_lf_m-\delta^i_k\delta^j_mf_l
-\delta^i_l\delta^j_kf_m+\delta^i_l\delta^j_mf_l
+\delta^i_m\delta^j_kf_l-\delta^i_m\delta^j_lf_k)\nonumber\\
\Delta_{c_i\bar c_j\bar c_k\bar c_l\bar c_m}=f^i_{jklm}+
\frac{1}{2} (\delta^i_jf_{klm}-\delta^i_kf_{jlm}+ 
\delta^i_lf_{jkm}-\delta^i_mf_{jkl} )\nonumber\\
\Delta_{c_ic_jc_kc_lc_m}=\epsilon^{ijklm} f,~~~
\Delta_{\bar c_i\bar c_j\bar c_k\bar c_l\bar c_m}=
\epsilon_{ijklm} \bar f
\end{eqnarray}
In terms of the irreducible tensors the vertex that enters in
Eq.(13) can be decomposed as follows
\begin{eqnarray}
\Gamma_{\mu}\Gamma_{\nu}\Gamma_{\lambda}\Gamma_{\rho}\Gamma_{\sigma}
\Delta_{\mu\nu\lambda\rho\sigma}=
(\epsilon_{ijklm} b_i b_j b_k b_lb_m f
+\epsilon^{ijklm} b_i^{\dagger} b_j^{\dagger}b_k^{\dagger}
b_l^{\dagger}b_m^{\dagger} \bar f )+~~~~~\nonumber\\
+(15 b_i^{\dagger} f^i
-20 b_i^{\dagger} b_n^{\dagger}b_n f^i 
+5 b_i^{\dagger} b_n^{\dagger}b_n b_m^{\dagger}b_m f^i
+ 15 b_i f_i
-20 b_n^{\dagger}b_n b_i f_i 
+5  b_n^{\dagger}b_n b_m^{\dagger}b_m b_i f_i)\nonumber\\
+10(b_i^{\dagger} b_j^{\dagger}b_k^{\dagger}f^{ijk}
-b_i^{\dagger} b_j^{\dagger}b_k^{\dagger}b_n^{\dagger}b_n f^{ijk}
  +b_ib_jb_k f_{ijk} - b_n^{\dagger}b_n b_ib_jb_k f_{ijk})\nonumber\\
+ (60 b_i^{\dagger} b_j^{\dagger}b_k  f^{ij}_k
-30 b_i^{\dagger} b_j^{\dagger}b_k b_n^{\dagger}b_n f^{ij}_k
+ 60 b_i^{\dagger} b_j b_k f^{i}_{jk}
 -30 b_n^{\dagger}b_n b_i^{\dagger} b_j b_k f^{i}_{jk})\nonumber\\
+ (5b_i^{\dagger} b_j^{\dagger}b_k^{\dagger}
b_l^{\dagger}b_m  f^{ijkl}_m
+ 5b_i^{\dagger} b_jb_kb_lb_m  f^{i}_{jklm})
+(10b_i^{\dagger} b_j^{\dagger}b_k^{\dagger}
b_lb_m  f^{ijk}_{lm}
+ 10b_i^{\dagger} b_j^{\dagger}b_kb_lb_m  f^{ij}_{klm})
 \end{eqnarray}
 The fields that appear above are not yet properly normalized.
 To normalize the fields we carry out a field redefinition 
 so that 
\begin{eqnarray}
f=\frac{2}{\sqrt {15}}h,~~~
f^i=\frac{4\sqrt 2}{\sqrt 5}h^i,~~f^{ijk}=\frac{\sqrt 2}{\sqrt {15}}
\epsilon^{ijklm}h_{lm}\nonumber\\
f^i_{jklm}= \frac{\sqrt 2}{\sqrt {15}}\epsilon_{jklmn}h^{(S)ni},~~
 f^i_{jk}=\frac{2\sqrt 2}{\sqrt {15}} h^i_{jk},~~
  f^{ijk}_{lm}=\frac{2}{\sqrt {15}} h^{ijk}_{lm},
 \end{eqnarray}
The  kinetic energy for the $\bar{126}$ plet field 
$-\partial_{\alpha}\phi_{\mu\nu\lambda\rho\sigma}
\partial^{\alpha}\phi_{\mu\nu\lambda\rho\sigma}^{\dagger}$
in terms of the normalized fields is then given by
 \begin{eqnarray}
L_{kin}^{(\bar{126})}=-(\partial_{\alpha} h\partial^{\alpha} h^{\dagger}
 + \partial_{\alpha} h^{i}\partial^{\alpha} h^{i\dagger}
+ \frac{1}{2}\partial_{\alpha} h_{ij}
\partial^{\alpha} h_{ij}^{\dagger}
+ \frac{1}{2}\partial_{\alpha} h^{(S)ij}
\partial^{\alpha} h^{(S)ij\dagger}\nonumber\\
+\frac{1}{2}\partial_{\alpha} h_{jk}^i
\partial^{\alpha} h_{jk}^{i\dagger} 
+\frac{1}{3!2}\partial_{\alpha} h^{ijk}_{lm}
\partial^{\alpha} h^{ijk\dagger}_{lm})
\end{eqnarray}
where the numerical factors are to account for the permutations.

\end{document}